\documentclass[aps,prl,superscriptaddress,showpacs,floatfix,twocolumn]{revtex4}

\usepackage{graphicx}   

\newcommand{\pperp}{\mbox{$p_\mathrm{T}\,$}}
\newcommand{\pperpg}{\mbox{$p_\mathrm{T}^{\gamma}\,$}}

\begin{document}

\title{Measurement of direct photon production 
       in $p+p$ collisions at $\sqrt{s}=200$ GeV}

\newcommand{\abilene}{Abilene Christian University, Abilene, TX 79699, USA}
\newcommand{\acadsin}{Institute of Physics, Academia Sinica, Taipei 11529, Taiwan}
\newcommand{\banaras}{Department of Physics, Banaras Hindu University, Varanasi 221005, India}
\newcommand{\barc}{Bhabha Atomic Research Centre, Bombay 400 085, India}
\newcommand{\bnl}{Brookhaven National Laboratory, Upton, NY 11973-5000, USA}
\newcommand{\caucr}{University of California - Riverside, Riverside, CA 92521, USA}
\newcommand{\ciae}{China Institute of Atomic Energy (CIAE), Beijing, People's Republic of China}
\newcommand{\cns}{Center for Nuclear Study, Graduate School of Science, University of Tokyo, 7-3-1 Hongo, Bunkyo, Tokyo 113-0033, Japan}
\newcommand{\colorado}{University of Colorado, Boulder, CO 80309, USA}
\newcommand{\columbia}{Columbia University, New York, NY 10027 and Nevis Laboratories, Irvington, NY 10533, USA}
\newcommand{\dapnia}{Dapnia, CEA Saclay, F-91191, Gif-sur-Yvette, France}
\newcommand{\debrecen}{Debrecen University, H-4010 Debrecen, Egyetem t{\'e}r 1, Hungary}
\newcommand{\elte}{ELTE, E{\"o}tv{\"o}s Lor{\'a}nd University, H - 1117 Budapest, P{\'a}zm{\'a}ny P. s. 1/A, Hungary}
\newcommand{\fsu}{Florida State University, Tallahassee, FL 32306, USA}
\newcommand{\gsu}{Georgia State University, Atlanta, GA 30303, USA}
\newcommand{\hiroshima}{Hiroshima University, Kagamiyama, Higashi-Hiroshima 739-8526, Japan}
\newcommand{\ihepprot}{IHEP Protvino, State Research Center of Russian Federation, Institute for High Energy Physics, Protvino, 142281, Russia}
\newcommand{\illuiuc}{University of Illinois at Urbana-Champaign, Urbana, IL 61801, USA}
\newcommand{\isu}{Iowa State University, Ames, IA 50011, USA}
\newcommand{\jinrdubna}{Joint Institute for Nuclear Research, 141980 Dubna, Moscow Region, Russia}
\newcommand{\kek}{KEK, High Energy Accelerator Research Organization, Tsukuba, Ibaraki 305-0801, Japan}
\newcommand{\kfki}{KFKI Research Institute for Particle and Nuclear Physics of the Hungarian Academy of Sciences (MTA KFKI RMKI), H-1525 Budapest 114, POBox 49, Budapest, Hungary}
\newcommand{\korea}{Korea University, Seoul, 136-701, Korea}
\newcommand{\kurchatov}{Russian Research Center ``Kurchatov Institute", Moscow, Russia}
\newcommand{\kyoto}{Kyoto University, Kyoto 606-8502, Japan}
\newcommand{\labllr}{Laboratoire Leprince-Ringuet, Ecole Polytechnique, CNRS-IN2P3, Route de Saclay, F-91128, Palaiseau, France}
\newcommand{\lawllnl}{Lawrence Livermore National Laboratory, Livermore, CA 94550, USA}
\newcommand{\losalamos}{Los Alamos National Laboratory, Los Alamos, NM 87545, USA}
\newcommand{\lpc}{LPC, Universit{\'e} Blaise Pascal, CNRS-IN2P3, Clermont-Fd, 63177 Aubiere Cedex, France}
\newcommand{\lund}{Department of Physics, Lund University, Box 118, SE-221 00 Lund, Sweden}
\newcommand{\muenster}{Institut f\"ur Kernphysik, University of Muenster, D-48149 Muenster, Germany}
\newcommand{\myongji}{Myongji University, Yongin, Kyonggido 449-728, Korea}
\newcommand{\nagasaki}{Nagasaki Institute of Applied Science, Nagasaki-shi, Nagasaki 851-0193, Japan}
\newcommand{\newmex}{University of New Mexico, Albuquerque, NM 87131, USA }
\newcommand{\nmsu}{New Mexico State University, Las Cruces, NM 88003, USA}
\newcommand{\ornl}{Oak Ridge National Laboratory, Oak Ridge, TN 37831, USA}
\newcommand{\orsay}{IPN-Orsay, Universite Paris Sud, CNRS-IN2P3, BP1, F-91406, Orsay, France}
\newcommand{\peking}{Peking University, Beijing, People's Republic of China}
\newcommand{\pnpi}{PNPI, Petersburg Nuclear Physics Institute, Gatchina, Leningrad region, 188300, Russia}
\newcommand{\riken}{RIKEN (The Institute of Physical and Chemical Research), Wako, Saitama 351-0198, JAPAN}
\newcommand{\rikjrbrc}{RIKEN BNL Research Center, Brookhaven National Laboratory, Upton, NY 11973-5000, USA}
\newcommand{\saopaulo}{Universidade de S{\~a}o Paulo, Instituto de F\'{\i}sica, Caixa Postal 66318, S{\~a}o Paulo CEP05315-970, Brazil}
\newcommand{\seoulnat}{System Electronics Laboratory, Seoul National University, Seoul, South Korea}
\newcommand{\stonybrkc}{Chemistry Department, Stony Brook University, Stony Brook, SUNY, NY 11794-3400, USA}
\newcommand{\stonycrkp}{Department of Physics and Astronomy, Stony Brook University, SUNY, Stony Brook, NY 11794, USA}
\newcommand{\subatech}{SUBATECH (Ecole des Mines de Nantes, CNRS-IN2P3, Universit{\'e} de Nantes) BP 20722 - 44307, Nantes, France}
\newcommand{\tenn}{University of Tennessee, Knoxville, TN 37996, USA}
\newcommand{\titech}{Department of Physics, Tokyo Institute of Technology, Oh-okayama, Meguro, Tokyo 152-8551, Japan}
\newcommand{\tsukuba}{Institute of Physics, University of Tsukuba, Tsukuba, Ibaraki 305, Japan}
\newcommand{\vandy}{Vanderbilt University, Nashville, TN 37235, USA}
\newcommand{\waseda}{Waseda University, Advanced Research Institute for Science and Engineering, 17 Kikui-cho, Shinjuku-ku, Tokyo 162-0044, Japan}
\newcommand{\weizmann}{Weizmann Institute, Rehovot 76100, Israel}
\newcommand{\yonsei}{Yonsei University, IPAP, Seoul 120-749, Korea}
\affiliation{\abilene}
\affiliation{\acadsin}
\affiliation{\banaras}
\affiliation{\barc}
\affiliation{\bnl}
\affiliation{\caucr}
\affiliation{\ciae}
\affiliation{\cns}
\affiliation{\colorado}
\affiliation{\columbia}
\affiliation{\dapnia}
\affiliation{\debrecen}
\affiliation{\elte}
\affiliation{\fsu}
\affiliation{\gsu}
\affiliation{\hiroshima}
\affiliation{\ihepprot}
\affiliation{\illuiuc}
\affiliation{\isu}
\affiliation{\jinrdubna}
\affiliation{\kek}
\affiliation{\kfki}
\affiliation{\korea}
\affiliation{\kurchatov}
\affiliation{\kyoto}
\affiliation{\labllr}
\affiliation{\lawllnl}
\affiliation{\losalamos}
\affiliation{\lpc}
\affiliation{\lund}
\affiliation{\muenster}
\affiliation{\myongji}
\affiliation{\nagasaki}
\affiliation{\newmex}
\affiliation{\nmsu}
\affiliation{\ornl}
\affiliation{\orsay}
\affiliation{\peking}
\affiliation{\pnpi}
\affiliation{\riken}
\affiliation{\rikjrbrc}
\affiliation{\saopaulo}
\affiliation{\seoulnat}
\affiliation{\stonybrkc}
\affiliation{\stonycrkp}
\affiliation{\subatech}
\affiliation{\tenn}
\affiliation{\titech}
\affiliation{\tsukuba}
\affiliation{\vandy}
\affiliation{\waseda}
\affiliation{\weizmann}
\affiliation{\yonsei}
\author{S.S.~Adler}	\affiliation{\bnl}
\author{S.~Afanasiev}	\affiliation{\jinrdubna}
\author{C.~Aidala}	\affiliation{\columbia}
\author{N.N.~Ajitanand}	\affiliation{\stonybrkc}
\author{Y.~Akiba}	\affiliation{\kek} \affiliation{\riken}
\author{A.~Al-Jamel}	\affiliation{\nmsu}
\author{J.~Alexander}	\affiliation{\stonybrkc}
\author{K.~Aoki}	\affiliation{\kyoto}
\author{L.~Aphecetche}	\affiliation{\subatech}
\author{R.~Armendariz}	\affiliation{\nmsu}
% ...
\collaboration{PHENIX Collaboration} \noaffiliation

\date{\today}

\begin{abstract}

Cross sections for midrapidity production of direct photons in
$p\!+\!p$ collisions at the Relativistic Heavy Ion Collider (RHIC) 
are reported for transverse momenta of $3\!<\!\pperp\!<\!16$ GeV/$c$.
Next-to-leading order (NLO) perturbative QCD (pQCD) describes 
the data well for $\pperp>5$ GeV/$c$, where the uncertainties of the 
measurement and theory are comparable.
We also report on the effect of requiring 
the photons to be isolated from parton jet energy.
The observed fraction of isolated photons is well described by
pQCD for $\pperp\!>\!7$ GeV/$c$.

\end{abstract}

        
\maketitle

The production of direct photons, i.e.~photons not from hadronic 
decays, in hadron-hadron collisions has been recognized as providing 
direct access to the gluon distributions in the hadron, both 
unpolarized and polarized \cite{Papavassiliou:1982qu,Berger:1989gz}. 
The process of direct photon production is described, at high energy 
and high momentum transfer, by perturbative Quantum Chromodynamics 
(pQCD). Three parton-parton subprocesses dominate at lowest order: 
Compton scattering $g+q\rightarrow \gamma +q$; annihilation 
$q+\bar{q}\rightarrow \gamma +g$; and parton-parton hard scattering 
with the scattered quark or gluon fragmenting to a photon. Where $g$ 
($q$) represent gluon (quark) states.  At next to leading order (NLO), 
bremstrahlung emission of photons from the quarks undergoing hard 
scattering also contributes to the direct photon signal.  The 
annihilation process is suppressed for $p+p$ collisions, due to the 
lower probability density of $\bar{q}$ versus $g$ in the proton.  In 
general, the fragmentation and bremstrahlung processes will produce 
photons in the vicinity of parton jets.  Therefore, a requirement that 
the photon be isolated from parton jet activity can emphasize the 
Compton graph.  Here, only the gluon distribution is unknown, 
particularly for the polarized case, and direct photon production 
therefore provides direct access to this (polarized) gluon 
distribution.

Comparisons of data to theory test our understanding of direct photon 
production in hadron-hadron collisions. Previous experiments have 
shown significant disagreement between data and theory at fixed target 
energy, and good agreement at collider energy 
\cite{Vogelsang:1997cq,Aurenche:1998gv}.  Results from the 
Relativistic Heavy Ion Collider (RHIC) for $p+p$ collisions cover 
intermediate energy and momentum transfer, overlapping CERN 
Intersecting Storage Rings (ISR) and Super antiProton Proton 
Synchrotron ($Sp\bar{p}S$) collider kinematics, and address the 
robustness of the pQCD prediction for direct photon production. In 
addition, the comparison of the direct photon rate using no isolation 
requirement, to the rate of observed photons that are isolated from 
parton jets, tests our understanding of the processes of parton 
fragmentation to photons, and of the bremstrahlung emission of photons 
from quarks in hard scattering.

Furthermore, direct photon production in $p+p$ collisions provides a
valuable baseline for the interpretation of direct photon data from
heavy-ion ($A+A$) collisions.  Jet-quenching models attribute the
strong suppression of high-\pperp hadrons in central $A+A$ collisions
to energy loss of scattered quarks and gluons in the hot and dense
medium created in these collisions~\cite{Adcox:2004mh}. Since photons
interact with the medium only electromagnetically, they provide a monitor
of the initial parton flux and therefore test a crucial assumption
of these models.

In this Letter, we present cross sections for direct photon production
in $p+p$ collisions at $\sqrt{s}\!=\!200$ GeV, from the 2003 run of
RHIC, at mid-rapidity for $3\!<\!\pperp\!<\!16$ GeV/$c$.
An earlier measurement \cite{Adler:2005qk} from the 2002 run of RHIC 
covered a much smaller region of \pperp.  Unpolarized cross sections 
are reported, obtained by averaging over the spin states of 
the beams, with $<\!1$\% residual polarization.

The data were collected by the PHENIX detector\cite{Adcox:2003zm}. The 
primary detector for this measurement is an electromagnetic 
calorimeter (EMCal), consisting of two subsystems, a six sector 
lead-scintillator (PbSc), and a two sector lead glass (PbGl) detector, 
each located 5~m radially from the beam line.  Each sector covers a 
range of $|\eta|\!<\!0.35$ in pseudo-rapidity and 22.5$^{\circ}$ in 
azimuth.  The EMCal has fine granularity. Each calorimeter tower 
covers $\Delta\eta\times\Delta\phi \sim 0.01 \times 0.01$, and a tower 
contains $\sim\!80$\% of the photon energy hitting the center of the 
tower. Two photons from $\pi^0 \rightarrow \gamma \gamma$ decays are 
clearly resolved up to a $\pi^0$ \pperp of 12~GeV/$c$, and a shower 
profile analysis extends the $\gamma/\pi^0$ discrimination to beyond 
20~GeV/$c$. The energy calibration of each tower is obtained from 
minimum-ionizing tracks and from the reconstructed $\pi^0$ mass. The 
uncertainty on the energy scale is less than 1.5\%.

Beam-beam counters (BBC) positioned at pseudo-rapidities $3.1\! <\! 
|\eta|\! <\! 3.9 $ provide a minimum bias (MB) trigger. Events with 
high \pperp photons are selected by a level-1 trigger that requires a 
minimum energy deposit of 1.4 GeV in an overlapping tile of $4\!\times 
\!4$ towers of the EMCal in coincidence with the MB trigger. The MB 
trigger cross section is $\sigma_\mathrm{BBC} = (23.0 \pm 2.2)$\,mb, 
about 50\% of $\sigma^{pp}_\mathrm{inel}$. The efficiency bias due to 
the MB trigger in the 2003 run, $\epsilon_\mathrm{bias} = 0.79 \pm 
0.02$, is determined from the ratio of the yield of high \pperp 
$\pi^0$ with and without the MB trigger. An integrated luminosity 
($\mathcal{L}$) of 240 $\rm{nb}^{-1}$ after a vertex cut of $\pm 30$ 
cm is used in this analysis.

The first step in the analysis is to cluster the hit towers. If there 
are two tower energy maxima and at least one lower-energy tower 
between them, the cluster is split into two, with the energy of each 
tower divided between the two clusters according to electromagnetic 
shower profiles associated with the clusters. Photons are identified 
by a shower profile cut that was calibrated using test beam data, 
identified electrons, and decay photons from identified $\pi^0$.  The 
cut rejects $\sim50$\% of hadrons depositing $E\!>\!3$ GeV in the 
EMCal and accepts $\sim98$\% of real photons.  The charged particle 
veto of the photon sample is based on tracks in drift chambers 2 m 
from the beamline, and hits in the pad chamber (PC3) immediately in 
front of the EMCal.  Loss of photons from conversions in material 
before the EMCal is estimated using a GEANT \cite{GEANT:W5013} 
simulation and confirmed by the observed fraction of identified 
$\pi^0$ photons vetoed.  The conversion correction is 3\% for the 
drift chamber veto and $\sim\!8$\% for the PC3 veto.  Remaining 
non-photon background, including converting neutral hadrons and albedo 
from the magnet yokes, is also estimated from the GEANT simulation at 
$\sim1$\%.

The experimental challenge in direct-photon measurements is the large 
photon background from decays of hadrons, primarily from $\pi^0 
\rightarrow \gamma \gamma$ ($\sim\!80$\% of the decays) and $\eta 
\rightarrow \gamma \gamma$ ($\sim\!15$\%).  We use two techniques 
described below to subtract the decay background: a $\pi^0$ tagging 
method and a cocktail subtraction method.

In the $\pi^0$ tagging method, a candidate photon is tagged as a 
$\pi^0$ decay photon if it forms a pair with another photon in the 
mass range $105\! < \!M_{\gamma\gamma} \!<\! 165$~MeV 
($M_{\pi^0}\pm$3$\sigma$), with $E_\gamma\!>\!150$~MeV. A fiducial 
region for direct photon candidates excludes 10 towers (0.1 radians) 
from the edges of the EMCal, while partner photons are accepted over 
the entire detector, to improve the probability of observing both 
decay photons from the $\pi^0$.

This method overestimates the yield of photons from $\pi^0$ decays, 
$\gamma_{\pi^0}$, due to combinatorial background. A \pperp dependent 
correction ($\sim 10$\%) is estimated from a fit to the $\pi^0$ 
sidebands, with $\pm3\%$ uncertainty. The yield of direct photons, 
$\gamma_\mathrm{dir}$, is obtained from the inclusive photon yield, 
$\gamma_\mathrm{incl}$, using the equation \begin{equation} 
\gamma_\mathrm{dir}=\gamma_\mathrm{incl}-(1+\delta^{\gamma}_{h/\pi^0})(1+R^\mathrm{miss}_{\pi^0})\gamma_{\pi^0}, 
\label{eqn:pi0tag} \end{equation} where $R^\mathrm{miss}_{\pi^0}$ is 
the correction for missing photon partners to the $\pi^0$; 
$(1+R^\mathrm{miss}_{\pi^0})\gamma_{\pi^0}$ represents the total 
contribution of photons from $\pi^0$ decays in each 
$p^\gamma_\mathrm{T}$ bin and $\delta^{\gamma}_{h/\pi^0}$ is the 
fraction of photons from hadrons other than $\pi^0$.

\begin{figure}[t]
\includegraphics[width=1.0\linewidth]{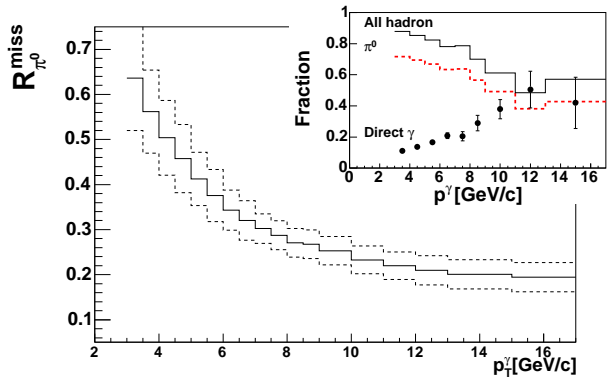}
\caption{\label{fig:pi0tagmethod}
Correction for missing photon partners to the $\pi^0$ 
($R^\mathrm{miss}_{\pi^0}$) vs. $p^\gamma_\mathrm{T}$. Dashed lines 
show the systematic uncertainty. Inset: Different contributions to the 
inclusive photon spectrum. Solid (dashed) lines represent all hadronic 
($\pi^0$) decay contributions. The data points show the remaining 
photon contributions.}
\end{figure}

To estimate $R^\mathrm{miss}_{\pi^0}$, a Monte Carlo simulation is 
used that includes the acceptance, energy resolution and our measured 
$\pi^0$ spectrum \cite{Adler:2003pb} as input. 
Figure~\ref{fig:pi0tagmethod} shows $(1+R^\mathrm{miss}_{\pi^0})$ from 
the simulation. The largest uncertainty is from the calibration of the 
EMCal at low energy. $\delta^{\gamma}_{h/\pi^0}$ is estimated by a 
simulation of hadron decays based on the 
$\eta/\pi^0$~\cite{Adler:2006hu} and 
$\omega/\pi^0$~\cite{Ryabov:2005xv} ratios from our measurements: 
$\delta^{\gamma}_{h/\pi^0}\approx0.24$ with 
$\delta^{\gamma}_{\eta/\pi^0}=0.19$ and 
$\delta^{\gamma}_{\omega/\pi^0}=0.05$. The contribution from other 
hadrons is less than 0.01. A small \pperp dependence is assumed to 
follow $m_\mathrm{T}$ scaling~\cite{Adcox:2002cg}. The inset of Fig.1 
shows the fraction of photons from $h$, $\pi^0$ and 
$\gamma_\mathrm{dir}$ to $\gamma_\mathrm{incl}$. The direct photon 
fraction ranges from 10\% at low $p_\mathrm{T}^{\gamma}$ to 50\% for 
$p_\mathrm{T}^{\gamma} > 10$ GeV.

In the cocktail method \cite{Adler:2005qk,Adler:2005ig}, the spectrum 
of decay photons is simulated utilizing our measured $\pi^0$ spectrum 
and applying $m_T$ scaling in order to account for other hadronic 
contributions. The effect of shower merging is also taken into account 
in the simulation. A double ratio $R_\gamma = 
(\gamma/\pi^0)^\mathrm{data}/(\gamma/\pi^0)^\mathrm{sim}$, is 
calculated for each \pperp bin. $R_\gamma > 1$ indicates a direct 
photon signal. The direct photon yield is extracted as 
$\gamma_\mathrm{dir} = (1 - R^{-1}_\gamma)\cdot \gamma_\mathrm{incl}$. 
Using the $\gamma/\pi^0$ ratio has the advantage that some systematic 
uncertainties cancel.

A summary of the systematic uncertainties is presented in Table 
\ref{table:syserrors}. Uncertainties of similar contributions are 
grouped together: global quantities (a), the inclusive photon yield 
(b) and the direct photon background (c-e). The categories (a)-(d) 
apply to both analysis methods. Category (a) includes the 
uncertainties of the energy scale, luminosity, and geometrical 
acceptance. The main contribution to category (b) is the uncertainty 
of the non-photon background estimation. The uncertainty of the 
charged particle veto is based on a study of the cluster vs. track 
matching in the EmCal and the tracking detectors.  The uncertainty in 
the neutral hadron contamination is estimated from identified charged 
hadrons. We assign the estimate of the albedo contribution as its 
uncertainty. Category (c) includes uncertainties of the correction for 
combinatorial background as estimated by different parameterizations 
of the background shape and the uncertainties of the $\pi^0$ 
reconstruction efficiency. Category (d) refers to the uncertainty of 
contributions from hadronic decays other than $\pi^0$'s, derived from 
our measurement of the hadron production ratios. Finally, category (e) 
combines all remaining uncertainties separately for the two analysis 
methods. Non-linearity effects in the energy calibration affect the 
minimum energy cut in the $\pi^0$ tagging method ($\rm{e}_1$) and 
distort the $\pi^0$ spectra in the cocktail method ($\rm{e}_2$).  
After the individual calibration a difference in the $\gamma/\pi^0$ 
ratio of PbGl and PbSc remains $(5-7\%)$. This is used to assign a 
systematic uncertainty of the non-linear part of the energy scale. Due 
to the small signal fraction at low \pperp, this translates into the 
large relative uncertainty in the direct photon spectra in Table 
\ref{table:syserrors}. In addition the uncertainty of the shower 
profile analysis of the $\gamma/\pi^0$ discrimination at high \pperp 
is included in this category. The two uncertainties 
($\rm{e}_1$,$\rm{e}_2$) are combined by averaging the squared 
uncertainties, and then all uncertainties were added in quadrature.

\begin{table}[b] 
\caption{
Relative systematic uncertainties of the direct photon spectra.
\label{table:syserrors}}
\begin{ruledtabular}
\begin{tabular}{l r r r}
\pperp[GeV/$c$] & 4.5-5 & 7.5-8 & 10-12 \\
Signal fraction & 9\% & 27\% & 49\% \\
\hline
(a) Global                 & 16.8\% & 14.9\% & 14.9\% \\
(b) Inclusive photons      & 12.3\% &  4.7\% &  3.1\% \\
(c) Photons from $\pi^0$   & 30.1\% & 10.7\% &  6.5\% \\
(d) Other hadrons          & 21.4\% &  6.7\% &  3.8\% \\
(e) Non-linearity (+ remaining) &        &        &         \\
\ \ ($\rm{e}_1$) $\pi^0$ tagging & 42.7\% &  6.8\% &  5.4\% \\
\ \ ($\rm{e}_2$) cocktail        & 69.5\% & 20.4\% & 13.4\% \\
\hline
Total                            & 71.6\% & 25.2\% & 19.8\% \\
\end{tabular}
\end{ruledtabular}
\end{table}

The results from the tagging and cocktail methods, obtained from 
independent analyses, agree within systematic uncertainties.  We 
report an average of the results, and uncertainties, of the two 
methods giving equal weight to the two analysis methods. 
average of the results, and uncertainties, of the 
approach gives equal weight to the two analysis methods.

\begin{figure}[tbh]
\includegraphics[width=1.0\linewidth]{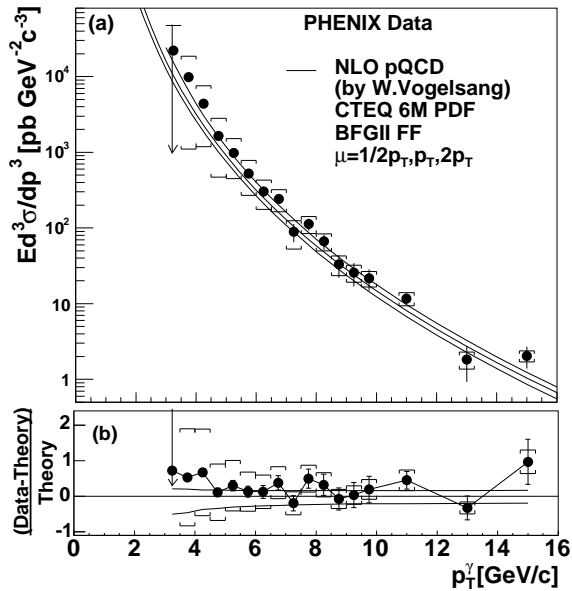}
\caption{\label{fig:spectra}
(a) Direct photon spectra with NLO pQCD calculations for 
three theory scales, $\mu$. 
Brackets around data points show systematic errors.
(b) Comparison to the NLO pQCD calculation for $\mu=\pperp$, 
with upper and lower curves for $\mu=\pperp/2$ and 2\pperp.}
\end{figure}

The invariant cross section of direct photon production is calculated 
by the following formula, \begin{equation} 
E\frac{d^3\sigma}{dp^3}=\frac{1}{\mathcal{L}} \frac{1}{2\pi \pperp} 
\frac{\gamma_\mathrm{dir}}{\Delta \pperp \Delta y} \frac{1}{\epsilon} 
\frac{1}{\epsilon_\mathrm{bias}}, \end{equation} where $\epsilon$ 
includes geometrical acceptance and the smearing effect from the 
energy resolution. The data points are plotted at the bin centers, 
with a correction to take into account the effect of finite bin sizes. 
The uncertainty of this correction is small compared to other 
systematic uncertainties.

Figure~\ref{fig:spectra} shows the measured invariant cross section 
for mid-rapidity direct photon production at $\sqrt{s}=200$~GeV.  In 
addition, a NLO pQCD prediction 
\cite{Gordon:1993qc,Gordon:1994ut,Aurenche:1983ws,Aurenche:1987fs,Baer:1990ra,Baer:1989xj}, 
using CTEQ~6M parton distribution functions \cite{Pumplin:2002vw} and 
the BFG~II parton to photon fragmentation function 
\cite{Bourhis:1997yu}, is shown with three theory scales ($\mu$) as 
indicated. The bottom panel shows the fractional difference between 
the data and this calculation.  The results are well described by 
pQCD.



The direct photon sample includes photons from the Compton and 
annihilation subprocesses, which are expected to be isolated from 
parton jet activity. To measure the fraction of isolated photons, we 
apply an isolation requirement in the $\pi^0$ tagging analysis method.  
Isolated photons are selected with less than 10\% additional energy 
within a cone of radius 
$\Delta r={\sqrt{(\Delta\eta)^2+(\Delta\phi)^2}}=$0.5 around the 
candidate photon direction. The cone energy is the sum of track 
momenta in the drift chamber and EMCal energy.  In most cases the cone 
is larger than the PHENIX acceptance and this is corrected for with a 
0.08 increase in the photon isolation fraction in the theory 
predictions below~\cite{Vogelsang:pc}.

\begin{figure}[tbh]
\includegraphics[width=1.0\linewidth]{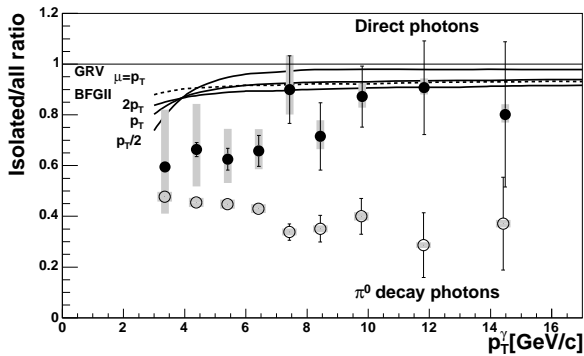}
\caption{\label{fig:isosubratio}
Closed circles: Ratio of isolated direct photons to all direct photons 
from the $\pi^0$-tagging method. The statistical uncertainties are 
shown as black error bars and the systematic uncertainties are plotted 
as shaded bars.  The solid and dashed curves are NLO pQCD calculations 
with three theory scales for BFGII \protect\cite{Bourhis:1997yu} and 
one scale for GRV \protect\cite{Gluck:1992zx} parton to photon 
fragmentation functions. Open circles: Ratio of isolated photons from 
$\pi^0$ decays to all photons from $\pi^0$ decays. }
\end{figure}

Figure~\ref{fig:isosubratio} presents the results of the isolation cut 
for photons from the $\pi^0$ tagging method.  Closed circles show the 
fraction of isolated direct photons to all direct photons. The curves 
are predictions from NLO pQCD, for the parton distribution and 
fragmentation functions as in Fig.~\ref{fig:spectra}, and for an 
additional parton to photon fragmentation function. The observed ratio 
is $\sim90\%$ for $\pperpg\!>\!7$ GeV/$c$ and it is well described by 
pQCD. An additional loss of $\sim15$ \%($\pperpg\!=\!3$ GeV/$c$) to 
less than 5 \%(for $\pperpg\!>\!10$ GeV/$c$) due to the underlying 
event is estimated by a PYTHIA\cite{Sjostrand:2003wg} simulation. 
Finally, for comparison, the open circles show the ratio of isolated 
photons from $\pi^0$ decays to all photons from $\pi^0$ decays. This 
indicates significantly less isolation than in the direct photon 
sample.

In summary, invariant cross sections for direct photon production at 
mid-rapidity have been measured up to $\pperp\!=\!16$ GeV/$c$ in 
$\sqrt{s}=200$ GeV $p+p$ collisions. The data are well described by 
NLO pQCD predictions for $\pperp\!>\!5$ GeV/$c$ where the 
uncertainties of the measurement and theory are comparable. When these 
data are combined with fixed target and Tevatron collider data, these 
measurements demonstrate the robustness of the pQCD description of 
direct photon production \cite{Aurenche:2006vj}.  In addition, the 
ratio of isolated photons to all non-hadronic decay photons is 
well-described by pQCD for $\pperp\!>\!7$ GeV/$c$.

Based on the comparison of high \pperp direct photon data from $Au+Au$ 
collisions at RHIC with a $p+p$ reference from NLO pQCD, the origin of 
the observed suppression of high-\pperp hadrons in central $Au+Au$ 
collisions can be attributed to properties of the hot and dense matter 
created in the $Au+Au$ collision \cite{Adler:2005ig}. The measurements 
presented here confirm this conclusion and put it on a firm 
experimental basis. Furthermore, the successful description of direct 
photon production at RHIC is a necessary test for the extraction of 
the gluon polarization from direct photon production in collisions of 
longitudinally polarized protons.



We thank the staff of the Collider-Accelerator and Physics
Departments at BNL for their vital contributions.
We thank Michel Fontannaz, Werner Vogelsang, and Monique Werlen 
for their interest and input.  We acknowledge support from
the Department of Energy and NSF (U.S.A.),
MEXT and JSPS (Japan),
CNPq and FAPESP (Brazil),
NSFC (China),
IN2P3/CNRS, CEA, and ARMINES (France),
BMBF, DAAD, and AvH (Germany),
OTKA (Hungary),
DAE and DST (India),
ISF (Israel),
KRF and CHEP (Korea),
RMIST, RAS, and RMAE (Russia),
VR and KAW (Sweden),
U.S. CRDF for the FSU,
US-Hungarian NSF-OTKA-MTA,
and US-Israel BSF.



\end{document}